\documentclass[aps, prb, letterpaper, 10pt, twocolumn, notitlepage, citeautoscript, groupedaddress, superscriptaddress, showpacs, english] {revtex4-2}

\usepackage[T1]{fontenc}
\usepackage[latin9]{inputenc}
\usepackage{babel}
\usepackage{amssymb, amsmath}
\usepackage{graphicx}
\usepackage{tikz}
\usetikzlibrary {arrows.meta}
\usepackage{bm}
\usepackage[colorlinks=true, linkcolor=black, citecolor=blue, urlcolor=blue, unicode=true]{hyperref}
\usepackage{xcolor}
\definecolor{darkyellow}{HTML}{aaaa00}
\definecolor{darkgreen}{HTML}{009900}
\definecolor{darkblue}{HTML}{0000ff}
\definecolor{turquois}{HTML}{00bfbf}


\begin{document}

\title{Magnetic Field Control of the N\'eel Vector and Magnon Visibility in Altermagnetic MnTe}

\newcommand{\ill}{Institut Laue-Langevin (ILL), 71 avenue des Martyrs, 38000 Grenoble, France}
\newcommand{\ess}{European Spallation Source (ESS), Data Management \& Scientific Computing Centre (DMSC), 2800 Lyngby, Denmark}
\newcommand{\usf}{Department of Physics, University of South Florida, Tampa, FL 33620, USA}
\newcommand{\um}{Department of Physics, University of Michigan, Ann Arbor, MI 48109, USA}
\newcommand{\caltech}{Department of Physics and Institute for Quantum Information and Matter, California Institute of Technology, Pasadena, CA 91125, USA}

\author{Tobias Weber}
\email[Correspondence: ]{tweber@ill.fr}
\affiliation{\ill}

\author{Niclas Heinsdorf}
\affiliation{\caltech}

\author{Shane Smolenski}
\affiliation{\um}

\author{Sawyer Beltz}
\affiliation{\usf}

\author{Alexandre Ivanov}
\affiliation{\ill}

\author{Andrea Piovano}
\affiliation{\ill}

\author{Victor Mecoli}
\affiliation{\ill}
\affiliation{\ess}

\author{Sean Knapp}
\affiliation{\usf}

\author{Ruiqi Tang}
\affiliation{\um}

\author{Amir Henderson}
\affiliation{\um}

\author{A. K. M. Ashiquzzaman Shawon}
\affiliation{\um}

\author{Taylor Pierce-James}
\affiliation{\um}

\author{Na Hyun Jo}
\affiliation{\um}

\author{Denis Karaiskaj}
\affiliation{\usf}

\begin{abstract}
Altermagnetic order gives rise to momentum-dependent spin splitting of electronic and magnonic excitations even in the absence of a net magnetization. 
Here, we investigate the magnetic field dependence of the spin-wave spectrum of altermagnetic $\alpha$-MnTe using inelastic neutron scattering and linear spin-wave theory.
An in-plane magnetic field continuously reorients the N\'eel vector by overcoming the weak crystalline anisotropy, while remaining small compared with the dominant exchange scale. 
We find that this reorientation leaves the magnon energies and line widths essentially unchanged, but strongly modifies the measured spectral intensity through the transverse-momentum projection. 
Our results demonstrate a clear separation between the soft orientational degree of freedom of the antiferromagnetic order and the robust exchange-dominated chiral magnon spectrum.
This combination establishes $\alpha$-MnTe as a platform for reconfigurable magnon coupling, in which external fields tune how excitations interact with polarized probes without substantially altering their frequency or coherence.
\end{abstract}


\maketitle

\section{Introduction}
The emerging field of altermagnetism describes a distinct class of compensated magnets in which crystallographic symmetry lifts the spin degeneracy conventionally associated with antiferromagnetic electronic bands, producing spin-split band structures without relying on relativistic spin-orbit coupling \cite{PhysRevX.12.031042, Smejkal2022, vsmejkal2022anomalous, PhysRevX.12.011028, PhysRevLett.126.127701, ma2021multifunctional, PhysRevX.12.021016, hayami2019momentum, hayami2020bottom, hayami2020spontaneous, naka2019spin, yuan2021prediction, reichlova2020macroscopic, vsmejkal2020crystal}. 
Traditionally, collinear magnets were divided into two classes: (i) ferromagnets, characterized by a parallel alignment of magnetic moments, finite net magnetization, and spin-split electronic and magnon bands; and (ii) antiferromagnets, characterized by an antiparallel alignment of magnetic moments, zero net magnetization, and spin-degenerate bands \cite{PhysRevX.12.031042, Smejkal2022, vsmejkal2022anomalous, PhysRevX.12.011028, PhysRevLett.126.127701}.
Altermagnets combine the antiparallel spin order and zero net magnetization of antiferromagnets with the time-reversal-symmetry-breaking and spin-split band structures characteristic of ferromagnets.

In conventional antiferromagnets, opposite-spin sub-lattices are connected by inversion or fractional-translation symmetry that give rise to a global Kramers' degeneracy. 
In altermagnets, on the other hand, the up and down sub-lattices are related by crystallographic rotations such that bands are spin-degenerate only at momenta that are left invariant under those rotations. 
As a consequence, the band structures of altermagnets form intricate networks of nodal lines and planes. 
As is well established for topological semimetals, such nodal structures can generate pronounced Berry curvature
and related quantum-geometric effects \cite{nodal1,nodal2,Wilde2021,nodal3,nodal4,nodal5},
providing an intuitive explanation for the Hall responses of altermagnets \cite{vsmejkal2020crystal, vsmejkal2022anomalous, am_geometry, Smolenski2025},
despite the absence of an appreciable net magnetization.
While the present work focuses on magnons and field control of the magnetic ground state,
the N\'eel-vector orientation is also crucial for the anomalous response \cite{McClarty2024}
and has recently been shown to be tunable by strain \cite{strain_ahe}.
The combination of spin-split band structures and zero net magnetization in altermagnets has immense application potential, ranging from spintronics and magnonics to quantum information processing \cite{jungwirth2016antiferromagnetic, baltz2018antiferromagnetic, shao2021spin, qin2023room, chen2023octupole,rhmg-j1fv,monkman}. 

In the limit of vanishing spin-orbit interactions, magnets are classified by spin space groups \cite{brinkman1966theory,litvin1974spin, litvin1977spin,ren2023enumeration,chen2023spin,xiao2023spin,jiang2023enumeration,yang2024symmetry}, 
which describe arrangements of local moments that are decoupled from the crystal lattice. 
Altermagnets are not allowed to have a center of inversion on any bond connecting opposite moments, 
and therefore place weaker restrictions on spin-orbit interactions compared to antiferromagnets \cite{moriya}. 
Since spin-orbit interactions induce a Dzyaloshinskii-Moriya exchange that destabilizes collinear order in favor of non-collinear magnetic ground states, promising material candidates are often sought among compounds containing light elements with weak intrinsic spin-orbit coupling. 
Guided by such symmetry-based considerations \cite{PhysRevX.12.031042, Smejkal2022}, momentum-resolved spectroscopy experiments have revealed spin-split electron and magnon bands in a growing number of altermagnets \cite{krempasky2024altermagnetic,zhu2024observation,chenli,faure2025altermagnetism,dale2026relativisticnonrelativisticspinsplitting,singhmagnons,ironfluoride}. 

Many proposed altermagnets are metallic, where signatures of magnon splitting are expected to be broadened or masked by short quasi-particle lifetimes and the itinerant electronic background \cite{singhmagnons}. 
In contrast, recent inelastic neutron-scattering and angle-resolved photoemission spectroscopy experiments on the semiconductor $\alpha$-MnTe resolved a sharp magnon spectrum exhibiting a spin splitting of about 2 meV along the $\left(\overline{1.33}\ 0\ l \right)$ branch \cite{Liu2024, Povarov2025, Osumi2024}. 
The chirality of these magnons was further shown to be switchable by a small external magnetic field \cite{Liu2026}. 

In this work, we investigate the effect of an external magnetic field on the magnon band structure
and their scattering intensities in altermagnetic $\alpha$-MnTe,
combining linear spin-wave theory \cite{Toth2015} with inelastic neutron-scattering measurements. 
We further discuss the implications of the robustness and tunability of these magnon modes for
the research of potential magnonic devices.

\section{Theory}

\begin{figure}
    \begin{centering}
    \begin{tikzpicture}
        \draw (0, 0) node [ inner sep = 0 ] { \includegraphics[width=0.30\textwidth]{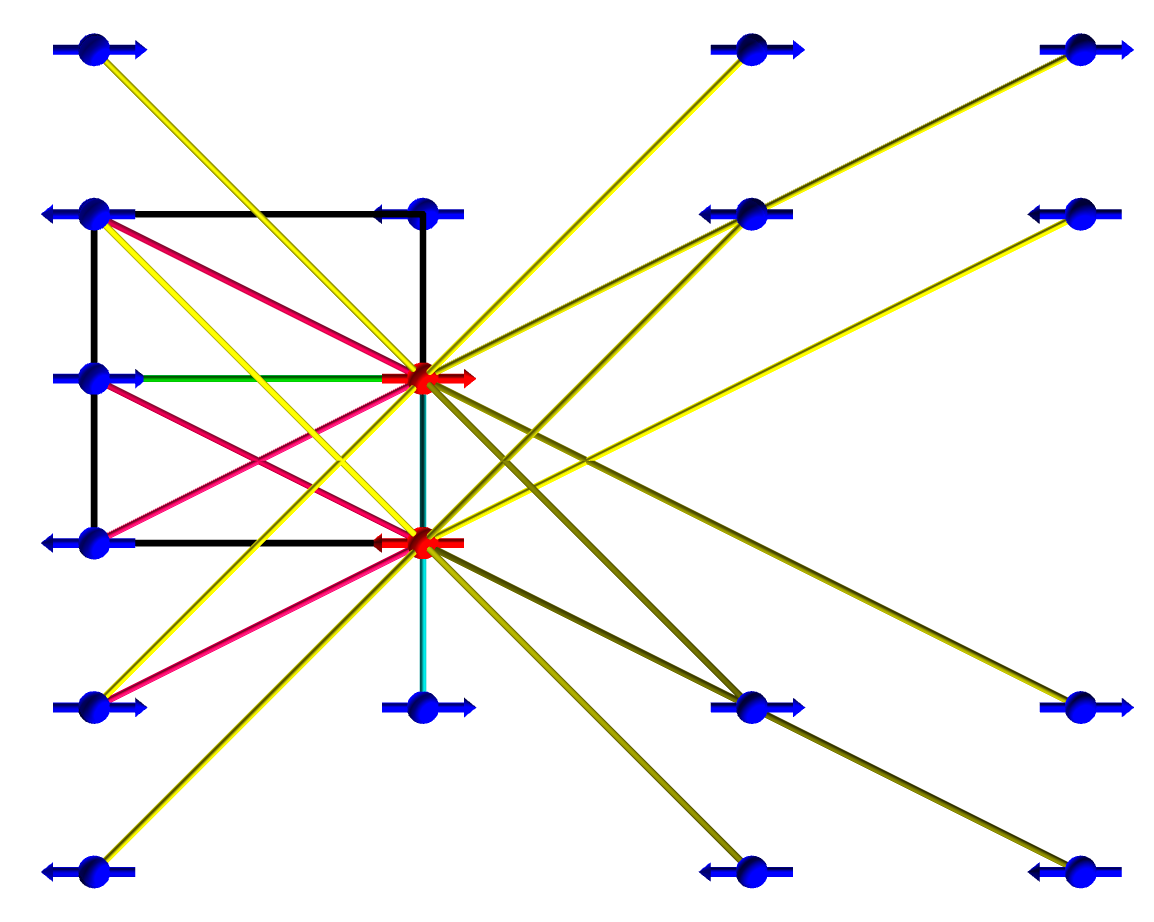} };
        \draw (2, 0) node [ inner sep = 0 ] { \includegraphics[width=0.035\textwidth]{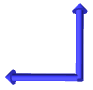} };
        \draw (4, 0) node [ inner sep = 0 ] { \includegraphics[width=0.15\textwidth]{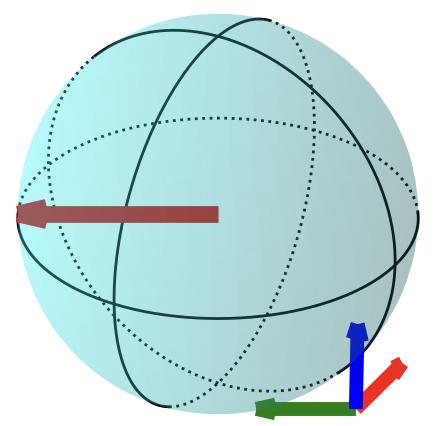} };
        \draw (-2.5, 2.25) node { \bf \fontfamily{phv} \selectfont (a) };
        \draw (4, -2) node { \fontfamily{phv} \selectfont B = 0 T };
        \draw (1.9, -0.45) node { \fontfamily{phv} \selectfont y };
        \draw (2.35, 0.05) node { \fontfamily{phv} \selectfont z };
        \draw (1.4, 1.2) node { \color{darkyellow}\fontfamily{phv} \selectfont $J_{10}$ };
        \draw (1.4, 0.45) node { \color{darkyellow}\fontfamily{phv} \selectfont $J_{11}$ };
        \draw (-1.7, 0.2) node { \color{darkgreen}\fontfamily{phv} \selectfont $J_2$ };
        \draw (-0.95, -0.8) node { \color{turquois}\fontfamily{phv} \selectfont $J_1$ };
        \draw (-1.7, -1.1) node { \color{red}\fontfamily{phv} \selectfont $J_3$ };
        \draw (3, 0.25) node { \bf \color{brown} \fontfamily{phv} \selectfont L };
        \draw (5.25, -0.9) node { \color{red} \fontfamily{phv} \selectfont x };
        \draw (4.25, -0.95) node { \color{darkgreen} \fontfamily{phv} \selectfont y };
        \draw (5, -0.75) node { \color{blue} \fontfamily{phv} \selectfont z };
    \end{tikzpicture}
    \vspace{0.2cm}
    \begin{tikzpicture}
        \draw (0, 0) node [ inner sep = 0 ] { \includegraphics[width=0.30\textwidth]{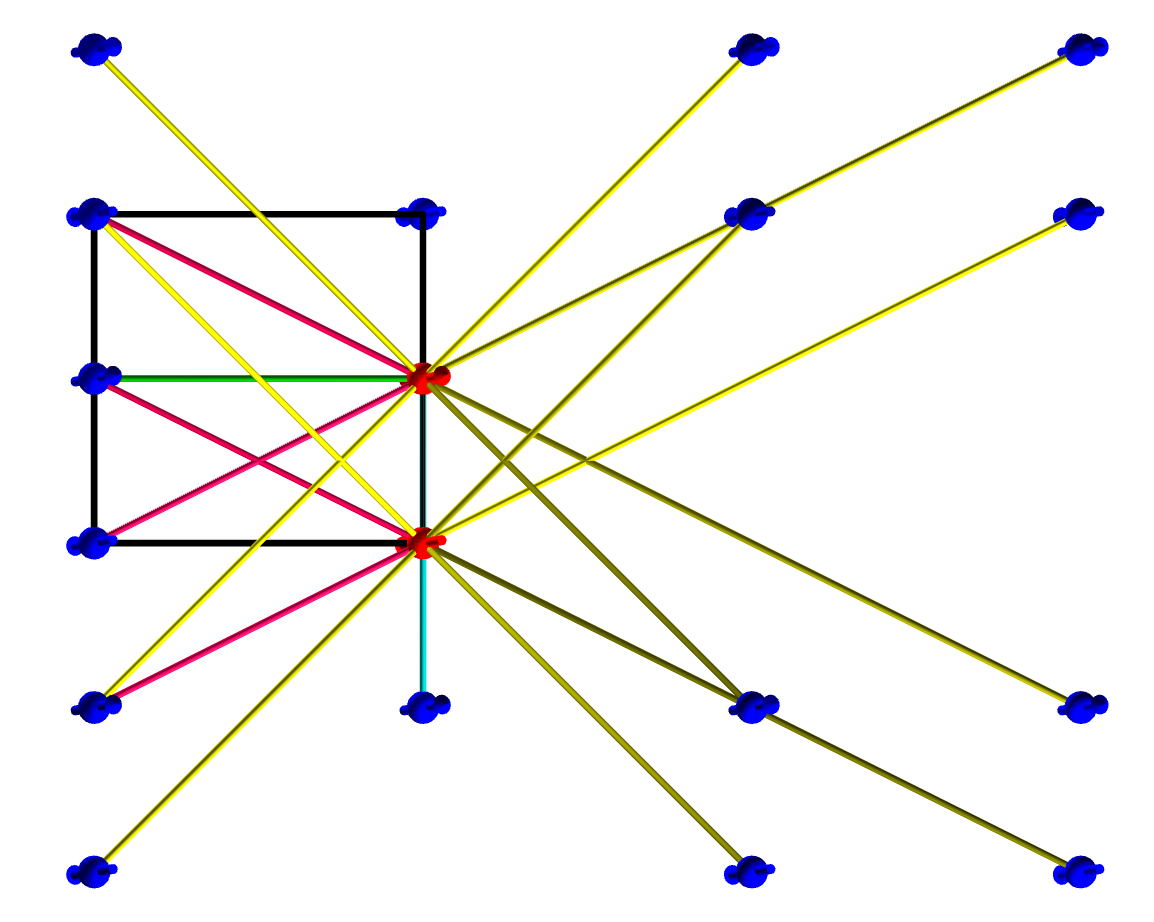} };
        \draw (2, 0) node [ inner sep = 0 ] { \includegraphics[width=0.035\textwidth]{figures/model_coords.png} };
        \draw (4, 0) node [ inner sep = 0 ] { \includegraphics[width=0.15\textwidth]{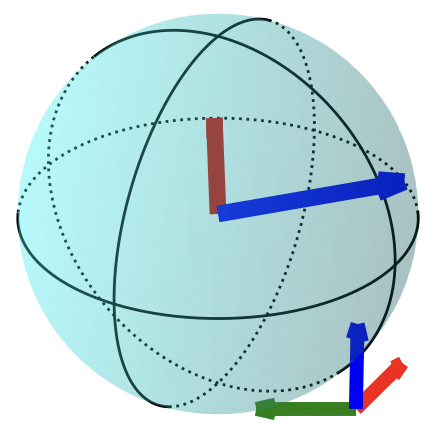} };
        \draw (-2.5, 2.25) node { \bf \fontfamily{phv} \selectfont (b) };
        \draw (4, -2) node { \fontfamily{phv} \selectfont B = 2.5 T };
        \draw (1.9, -0.45) node { \fontfamily{phv} \selectfont y };
        \draw (2.35, 0.05) node { \fontfamily{phv} \selectfont z };
        \draw (3.95, 0.85) node { \bf \color{brown} \fontfamily{phv} \selectfont L };
        \draw (5, 0.45) node { \bf \color{blue} \fontfamily{phv} \selectfont B };
        \draw (5.25, -0.9) node { \color{red} \fontfamily{phv} \selectfont x };
        \draw (4.25, -0.95) node { \color{darkgreen} \fontfamily{phv} \selectfont y };
        \draw (5, -0.75) node { \color{blue} \fontfamily{phv} \selectfont z };
    \end{tikzpicture}
    \vspace{0.2cm}
    \begin{tikzpicture}
        \draw (-0.3, 0) node [ inner sep = 0 ] { \includegraphics[width=0.25\textwidth]{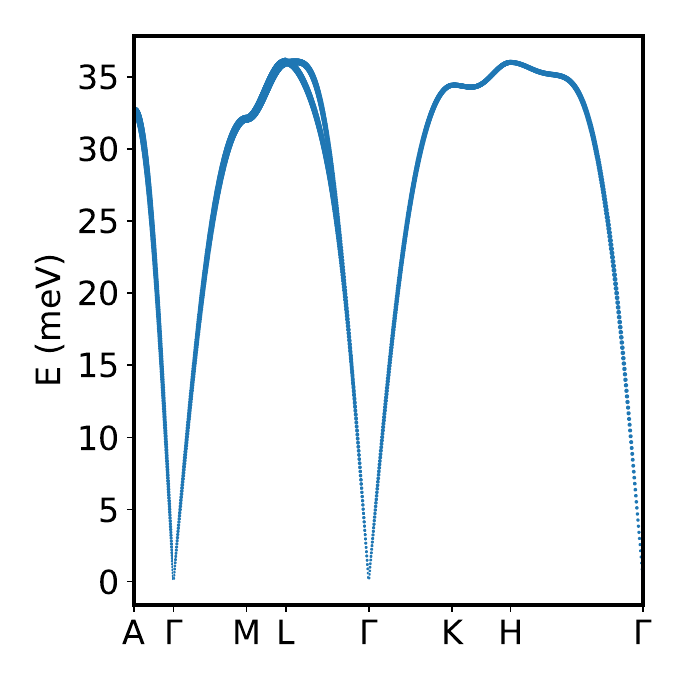} };
        \draw (3.8, 0) node [ inner sep = 0 ] { \includegraphics[width=0.2\textwidth]{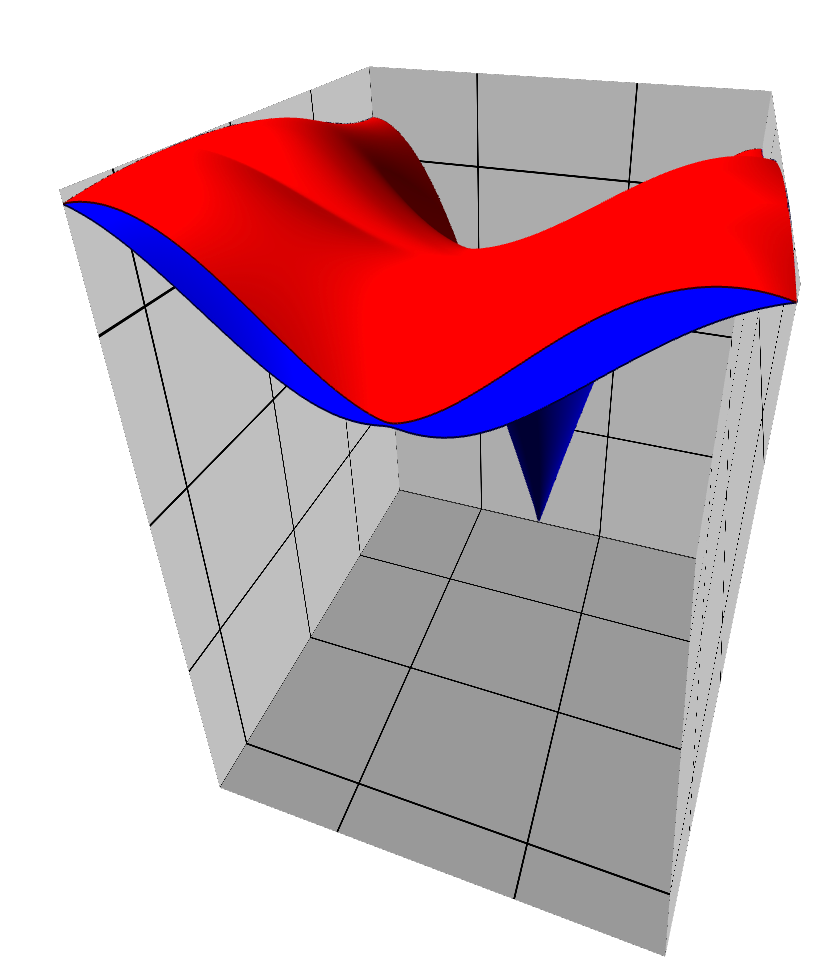} };

        \draw (3.4, -2.2) node { \fontfamily{phv} \selectfont $\left( 00l \right)$ };
        \draw (4.1, -2.) node { \fontfamily{phv} \selectfont 0.8 };
        \draw (3.4, -1.8) node { \fontfamily{phv} \selectfont 1.2 };

        \draw (2.4, 1.95) node { \fontfamily{phv} \selectfont $\left( h00 \right)$ };
        \draw (3.1, 1.9) node { \fontfamily{phv} \selectfont 1.2 };
        \draw (2.6, 1.7) node { \fontfamily{phv} \selectfont 1.4 };
        \draw (2.1, 1.5) node { \fontfamily{phv} \selectfont 1.6 };

        \draw (2.7, 0.15) node { \fontfamily{phv} \selectfont $E \left( \mathrm{meV} \right)$ };
        \draw (2.5, -0.9) node { \fontfamily{phv} \selectfont 10 };
        \draw (2.3, -0.3) node { \fontfamily{phv} \selectfont 20 };
        \draw (2.1, 0.6) node { \fontfamily{phv} \selectfont 30 };

        \draw (-2, 2.0) node { \bf \fontfamily{phv} \selectfont (c) };
        \draw (1.8, 2.0) node { \bf \fontfamily{phv} \selectfont (d) };
    \end{tikzpicture}
    \caption{(a) and (b) Magnetic model for MnTe \cite{Liu2024} showing only the magnetically active Mn sites in fractional coordinates (blue and red spheres)
    as well as the spin orientation (blue and red arrows) for (a) $B = 0\ \mathrm{T}$ and (b) $B = 2.5\ \mathrm{T}$,
    viewing along the $\left[ 100 \right]$ direction.
    The inset shows the orientation of the N\'eel vector relative to the magnetic field in more detail.
    Antiferromagnetic couplings, $J_1$, $J_3$, are shown in red and turquoise, ferromagnetic couplings, $J_2$, in green, and
    long-range altermagnetic couplings, $J_{10}$ and $J_{11}$, in light and dark yellow.
    The black rectangular area marks the boundaries of the nuclear unit cell.
    (c) Calculated magnon band structure along high-symmetry directions for $B=0\ \mathrm{T}$.
    (d) Calculated spin-wave dispersion along the experimentally accessible $\left(h0l\right)$ scattering plane 
    showing the spin-split magnon surfaces in red and blue.}
    \label{fig:model}
    \end{centering}
\end{figure}

$\alpha$-MnTe crystallizes in the centrosymmetric, hexagonal $\mathrm{P6_3/mmc}$ space group (SG $\#194$) \cite{Minikayev2015, Liu2024}. 
Below the N\'eel temperature of $T_N = 310\ \mathrm{K}$ \cite{Kriegner2017}, the compound exhibits antiferromagnetic order that is shown schematically in Fig.~\ref{fig:model}(a). 
The Mn magnetic moments align ferromagnetically within the $\left[ab\right]$-plane along the $\left[210\right]$ direction and antiferromagnetically between adjacent layers along $c$, 
stacked along a nearest-neighbor Mn--Mn bond \cite{Kriegner2017, Liu2024}. 
The Mn atoms occupy the $2a$ Wyckoff position, the site-symmetry group of which includes inversion symmetry. 
Its orbit is generated by a sixfold screw rotation resulting in an altermagnetic $g$-wave pattern \cite{Liu2024}.

We adopt the Heisenberg Hamiltonian from Ref. \cite{Liu2024},
\begin{align} \label{eq:hamiltonian}
	H_0 = &\sum_{\langle ij\rangle}
	J_{1}\hat{\mathbf{S}}_i \cdot \hat{\mathbf{S}}_j
	+  \sum_{\langle\langle ij\rangle\rangle}
	J_{2}\hat{\mathbf{S}}_i \cdot \hat{\mathbf{S}}_j
	+   \sum_{\langle\langle\langle ij\rangle\rangle\rangle}
	J_{3}\hat{\mathbf{S}}_i \cdot \hat{\mathbf{S}}_j \nonumber \\
    &\sum_{\langle ij\rangle_{10}}J_{10}\hat{\mathbf{S}}_i \cdot \hat{\mathbf{S}}_j
	+\sum_{\langle ij\rangle_{11}} J_{11}\hat{\mathbf{S}}_i \cdot \hat{\mathbf{S}}_j
	+  \sum_{i} A S{^z_i}^2,
\end{align} 
with exchange constants $J_1 = 3.993$ meV, $J_2 = -0.1202$ meV, $J_3 = 0.4723$ meV, $J_{10} = 0.06817$ meV, $J_{11} = -0.02212$ and single-ion anisotropy $A = 0.04825$ meV \cite{Liu2024}.
The exchange paths are depicted in Fig.~\ref{fig:model}(a). 
The effect of the external magnetic field $\mathbf{B}$ is modeled by adding a Zeeman term,
\begin{align}
    H_Z = g_e \mu_B \mathbf{B} \cdot \mathbf{S}_i.
\end{align}
The 10th- and 11th-nearest-neighbor bonds have the same length but are not related by any space-group symmetry; 
therefore, in general, $J_{10}\neq J_{11}$. 
These bonds constitute the shortest exchange paths on the Mn sub-lattice that break inversion symmetry between the spin-up and spin-down sites, thereby rendering the two-atom magnetic unit cell primitive.

The exchange interactions in Eq.~\eqref{eq:hamiltonian} favor translationally-invariant, compensated, collinear spin configurations. 
For a single-domain state, these are characterized by a uniform N\'eel vector $\mathbf{L}$, defined as the difference between sub-lattice magnetizations. 
At zero field, the anisotropy term spontaneously selects $\mathbf{L}$ along any of the three basal directions of the hexagonal lattice. 
Together with the two possible signs of $\mathbf{L}$, the threefold rotational symmetry about the $c$ axis yields six degenerate classical ground states:
$\mathbf{L} = \pm \left( 010 \right)$, $\pm \left( 100 \right)$, and $\pm \left( 1\bar{1}0 \right)$.
One is shown in panel (a) of Fig.~\ref{fig:model}. 

\begin{table}
\begin{centering}
\begin{tabular}{p{1cm} | p{2.5cm} | p{2.5cm} | p{1cm}}
	$B$ & $\left[L_x L_y L_z\right]$ & $\left(L_h L_k L_l\right)$ & $\measuredangle \left( \mathbf{B}, \mathbf{L} \right)$ \\
	\hline
	0 T     & $\left[ 0, 1, 0 \right]$               & $\left( 0, 1, 0 \right)$ & $150^\circ{}$   \\
	0.5 T  & $\left[ 0.368, 0.930, 0 \right]$ & $\left( 0.240, 0.405, 0 \right)$ & $128.4^\circ{}$ \\
	1 T     & $\left[ 0.762, 0.647, 0 \right]$ & $\left( 0.497, 0.117, 0 \right)$ & $100.3^\circ{}$ \\
	1.5 T  & $\left[ 0.825, 0.564, 0 \right]$ & $\left( 1.000, 0.091, 0 \right)$ & $94.4^\circ{}$ \\
	2 T     & $\left[ 0.843, 0.537, 0 \right]$ & $\left( 1.000, 0.052, 0 \right)$ & $92.5^\circ{}$ \\
	2.5 T  & $\left[ 0.851, 0.525, 0 \right]$ & $\left( 1.000, 0.034, 0 \right)$ & $91.7^\circ{}$ \\
	$\ge$ 5 T  & $\left[ 0.859, 0.513, 0 \right]$ & $\left( 1.000, 0.017, 0 \right)$ & $90.8^\circ{}$
\end{tabular}
\end{centering}
\caption{Calculated ground state N\'eel vector orientation $\left(L_h,\ L_k,\ L_l \right)$ shown for the
$\mathbf{L}_{B=0} = \left( 010 \right)$ domain and for different field magnitudes $B$ of an external field
along the $\left(1\bar{2}0\right)$ direction.
The second column shows the N\'eel vector in the absolute lab system, the third column in reciprocal coordinates.
Also note that for the $0\ \mathrm{T}$ case the field angle corresponds to the orientation of the coils, here, the field itself was turned off.}
\label{tab:groundstate}
\end{table}

Restricting the classical ground state to collinear configurations, an external magnetic field $\mathbf{B}$ reduces this sixfold degeneracy to twofold. 
The selected orientation depends on the relative strengths of $\mathbf{B}$ and the anisotropy $A$, while the much larger, rotationally invariant exchange energy does not favor a particular direction of $\mathbf{L}$.
In the spin-rotationally invariant limit, the system is agnostic to the orientation of $\mathbf{L}$.
However, even weak spin-lattice coupling makes the magnetic point group dependent on the N\'eel-vector direction \cite{pgs1}.
This symmetry change can allow or forbid a small ferromagnetic moment \cite{pgs2},
with important consequences for field-driven domain control \cite{Liu2026b},
strain coupling \cite{McClarty2024,pgs3,Kluczyk2024,strain_ahe},
and the anomalous Hall response \cite{vsmejkal2020crystal,vsmejkal2022anomalous,am_geometry,Smolenski2025}.

Starting from $\mathbf{L} = \left( 010 \right)$ at $B = 0$,
we use linear spin-wave theory to minimize the ground state energy as a function of field strength along
$\left( 1 \bar{2} 0 \right)$.
For numerical stability we add a tiny, negative single-ion anisotropy term of $-1.75\cdot10^{-5}\ \mathrm{meV}$ along the initial $\mathbf{L}$. 
The resulting N\'eel vectors are listed in Tab.~\ref{tab:groundstate}. 
Its rightmost column shows the angle between the $\mathbf{B}$ and $\mathbf{L}$, which saturates to 90$^\circ$
at approximately $1.5$ T and above.
The corresponding spin configuration is plotted in Fig.~\ref{fig:model}(b). 
Having demonstrated field control of the N\'eel vector, we next investigate the ground state excitations, and in particular the chiral magnons characteristic of altermagnetism.

The primitive unit cell contains two magnetic ions, giving rise to two magnon branches. 
Fig.~\ref{fig:model}(c, d) shows the linear-spin-wave spectrum for momentum transfers in the $\left(h0l\right)$ plane. 
The two branches are expected to be degenerate at momenta invariant under a crystallographic symmetry that exchanges the two Mn sites. 
Accordingly, they are twofold degenerate in the $\left[ 1\bar{1}0 \right]$ and $\left[ 001 \right]$ mirror planes. 
Along $\Gamma$--$L$, however, the branches are nondegenerate and, in the absence of spin-orbit coupling, fully spin-polarized. 
We refer to these excitations as chiral magnons, which are the fundamental carriers of information in magnonic devices \cite{chumak2015magnon}. 

Because magnons carry magnetic moments, an external field shifts their frequencies and can modify their lifetimes and nonlinear dynamics, even when their group velocities remain unchanged. 
It is therefore important to determine whether the field used to control the N\'eel vector interferes with the generation of coherent spin waves. 
Since the exchange scale greatly exceeds the Zeeman energy, we expect secondary effects associated with the field-induced rotation of $\mathbf{L}$ to dominate. 
Inelastic neutron scattering can probe both magnon damping through the excitation line shapes and the orientation of $\mathbf{L}$ through the transverse dynamic structure factor \cite[p. 47]{Shirane2002},
\begin{equation}
    S_{x y}^{\perp}\left(\mathbf{Q}, E\right) = 
        \left( \delta_{x y} - Q_{x} Q_{y} / Q^2 \right)
        \cdot S_{x y}\left(\mathbf{Q}, E\right),
    \label{equ:Sperp}
\end{equation}
where $x$ and $y$ are the cartesian coordinate indices.
The transverse projector selects correlations perpendicular to the momentum transfer $\mathbf{Q}$ and is therefore sensitive to the field-induced rotation of $\mathbf{L}$, while changes in lifetime and quasi-particle weight appear through the line shape and intensity of $S_{x y}(\mathbf{Q},E)$.

We next present inelastic neutron-scattering results that establish field control of the ordered ground state and characterize the field dependence of the magnon excitations.


\section{Experiments}
\begin{figure}
    \begin{centering}
    \begin{tikzpicture}
        \draw (0, 0) node [ inner sep = 0 ] { \includegraphics[width=0.66\columnwidth]{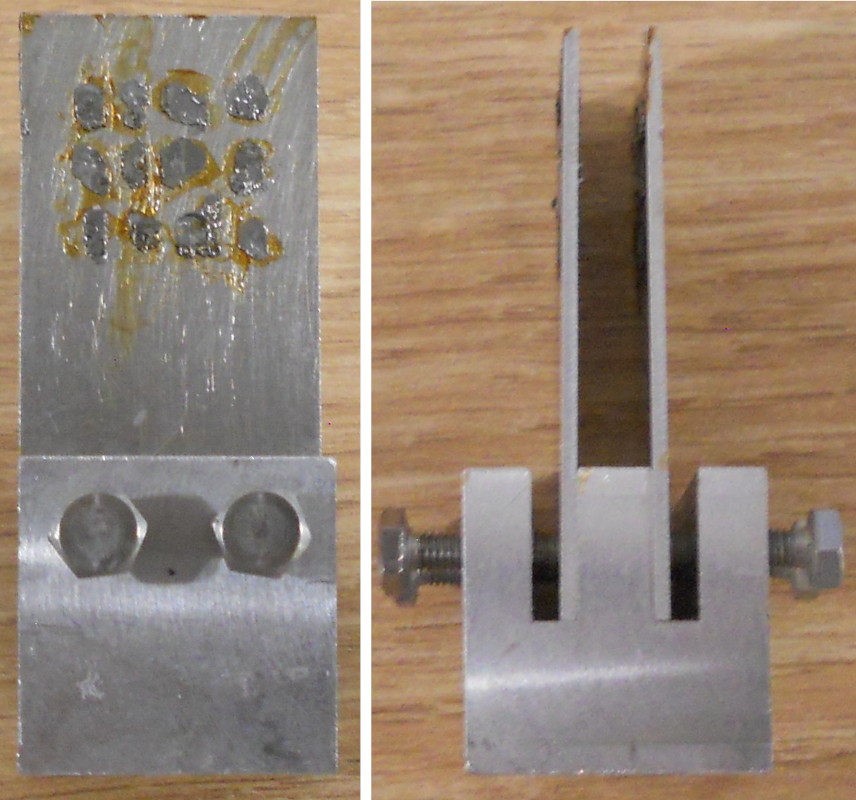} };

        \draw [line width = 1.75, black, arrows = {-Stealth}] (-2, -2.25) -- (-1, -2.25);
        \draw [line width = 1.75, black, arrows = {-Stealth}] (-2, -2.25) -- (-2, -1.25);
        \draw (-1, -2) node { \bf \color{black} \fontfamily{phv} \selectfont $\bm{\left[210\right]}$ };
        \draw (-1.75, -1.5) node { \bf \color{black} \fontfamily{phv} \selectfont b };

        \draw [line width = 1.75, black, arrows = {-Stealth}] (1, -2.25) -- (2, -2.25);
        \draw [line width = 1.75, black, arrows = {-Stealth}] (1, -2.25) -- (1, -1.25);
        \draw (2, -2) node { \bf \color{black} \fontfamily{phv} \selectfont c };
        \draw (1.25, -1.5) node { \bf \color{black} \fontfamily{phv} \selectfont b };
    \end{tikzpicture}
    \caption{
    The co-aligned sample used for inelastic neutron scattering. 
    It consists of the small MnTe single crystals glued to the front and back sides of two aluminum slabs.
    Left: front view, right: side view.}
    \label{fig:sample}
    \end{centering}
\end{figure}
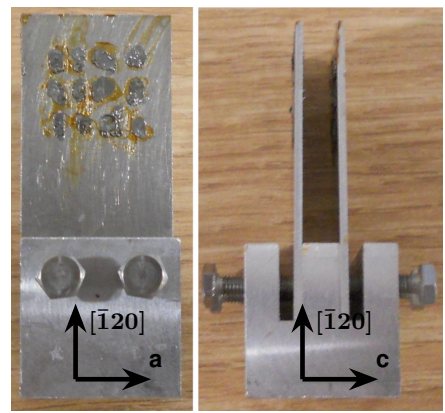

MnTe single crystals were grown by solution growth method out of Sb flux. 
An initial stoichiometry of Mn$_{7.3}$Te$_{14.5}$Sb$_{78.2}$ was placed in a fritted alumina crucible \cite{CANFIELD2016, SLADE2022} and sealed in a quartz ampoule under partial pressure of Ar. 
The ampoule was heated to $950^\circ{} \mathrm{C}$ over 5 hours, held there for 5 hours, and then cooled to $650^\circ{} \mathrm{C}$ over 60 hours. 
At this temperature, the excess solution was separated from the MnTe crystals using a centrifuge. 
The crystals form in mm-sized hexagonal plates and the structure was confirmed via powder X-ray diffraction with minimal contributions from residual surface Sb and Te flux.

\begin{figure*}[htb]
    \begin{centering}
    \begin{tikzpicture}
        \draw (0, 0) node [ inner sep = 0 ] { \includegraphics[width=0.31\textwidth]{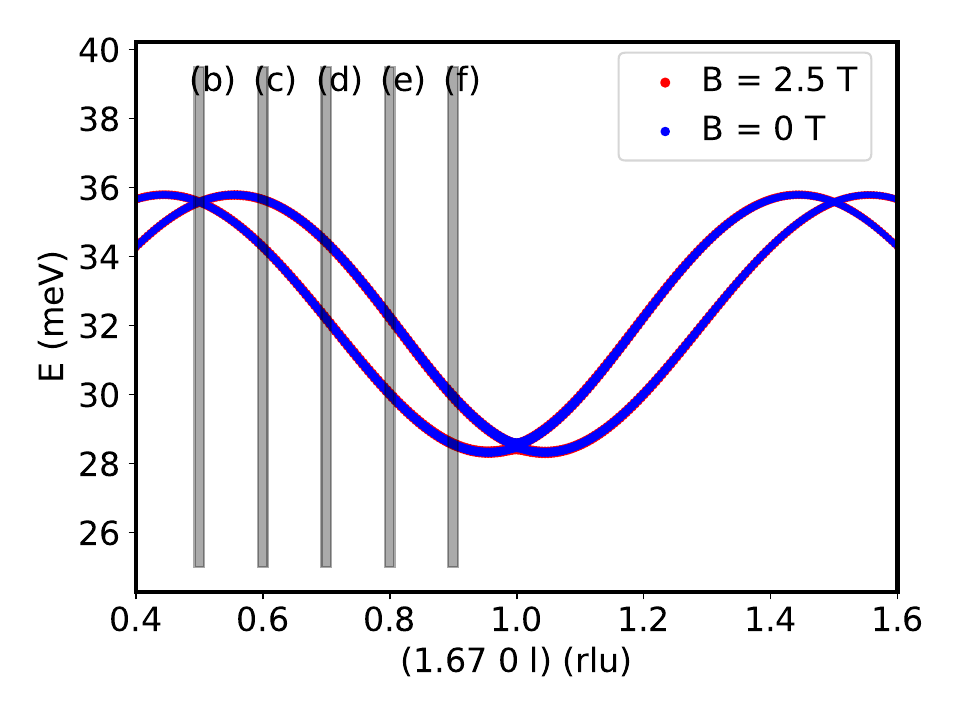} };
        \draw (-2.2, 2.2) node { \bf \fontfamily{phv} \selectfont (a) };
    \end{tikzpicture}
    \begin{tikzpicture}
        \draw (0, 0) node [ inner sep = 0 ] { \includegraphics[width=0.31\textwidth]{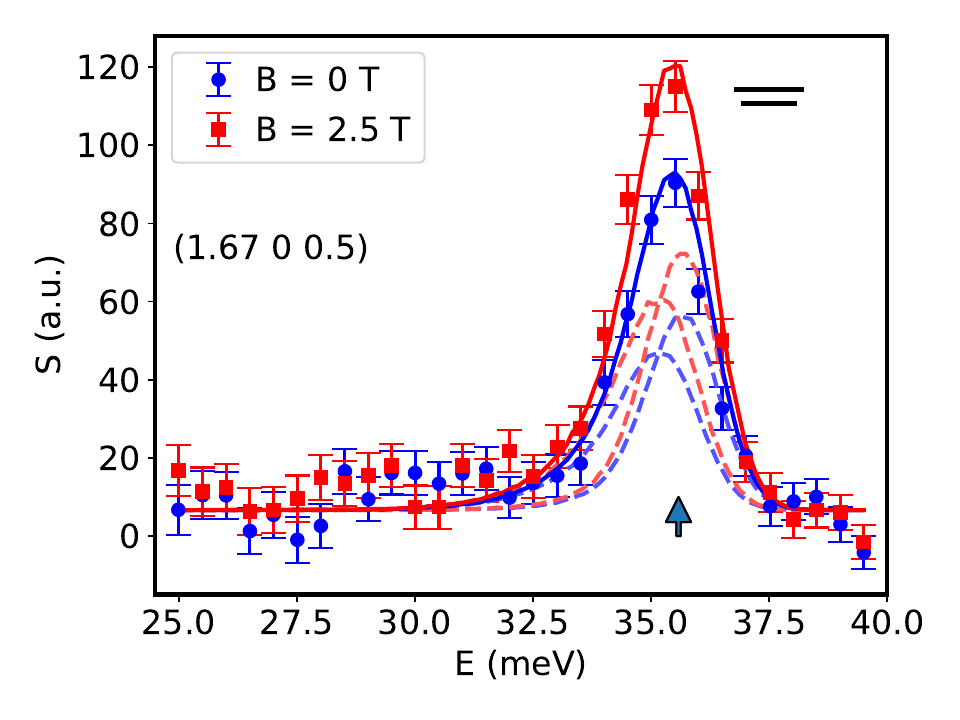} };
        \draw (-2.2, 2.2) node { \bf \fontfamily{phv} \selectfont (b) };
    \end{tikzpicture}
    \begin{tikzpicture}
        \draw (0, 0) node [ inner sep = 0 ] { \includegraphics[width=0.31\textwidth]{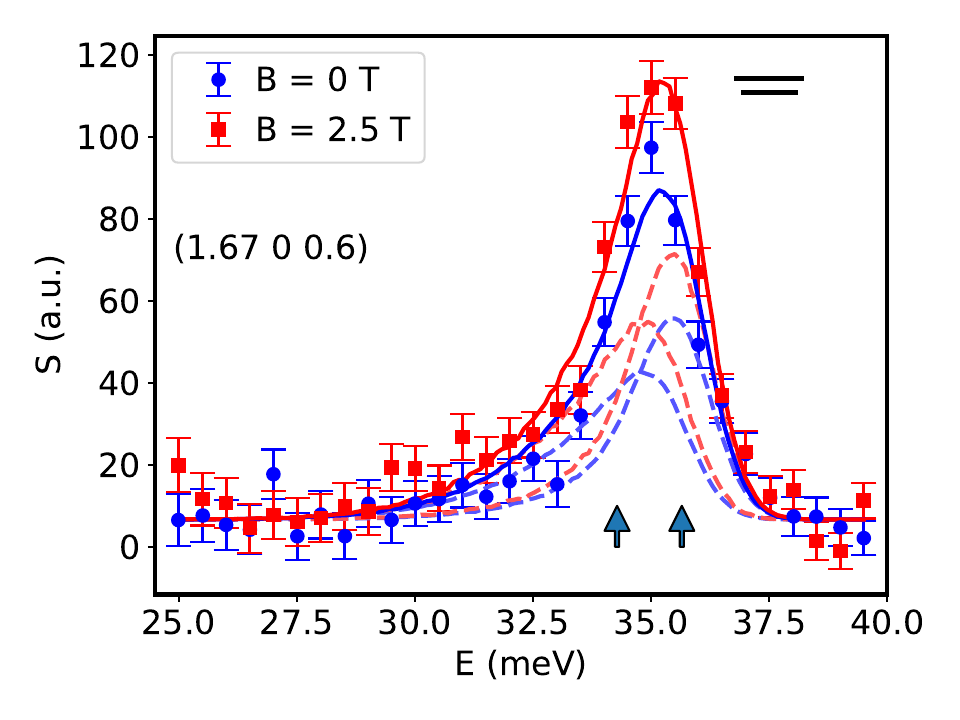} };
        \draw (-2.2, 2.2) node { \bf \fontfamily{phv} \selectfont (c) };
    \end{tikzpicture}
    \begin{tikzpicture}
        \draw (0, 0) node [ inner sep = 0 ] { \includegraphics[width=0.31\textwidth]{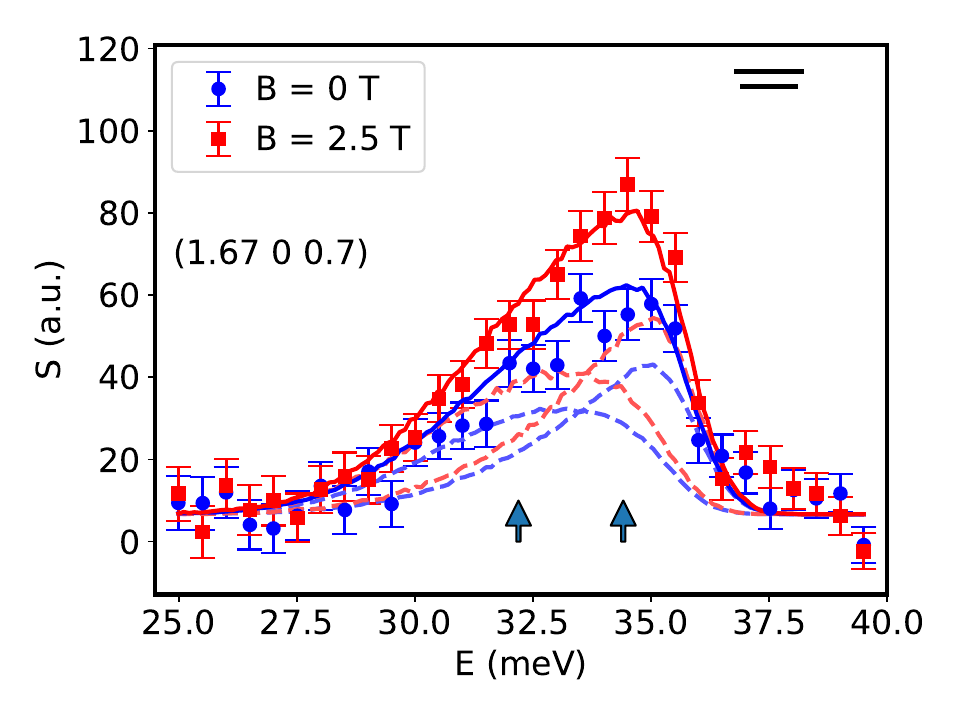} };
        \draw (-2.2, 2.2) node { \bf \fontfamily{phv} \selectfont (d) };
    \end{tikzpicture}
    \begin{tikzpicture}
        \draw (0, 0) node [ inner sep = 0 ] { \includegraphics[width=0.31\textwidth]{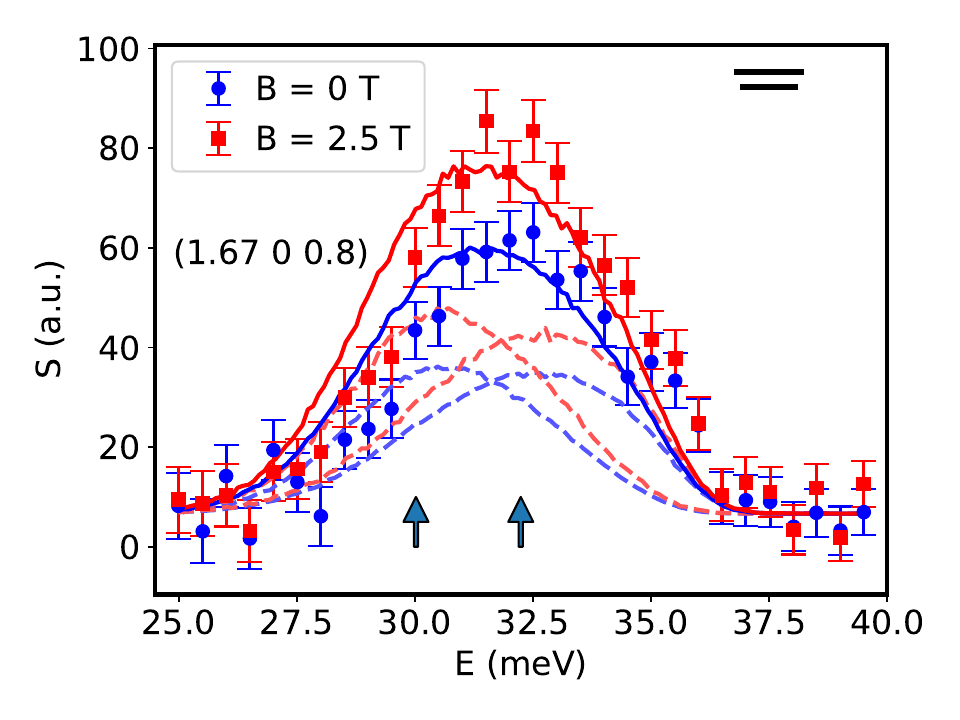} };
        \draw (-2.2, 2.2) node { \bf \fontfamily{phv} \selectfont (e) };
    \end{tikzpicture}
    \begin{tikzpicture}
        \draw (0, 0) node [ inner sep = 0 ] { \includegraphics[width=0.31\textwidth]{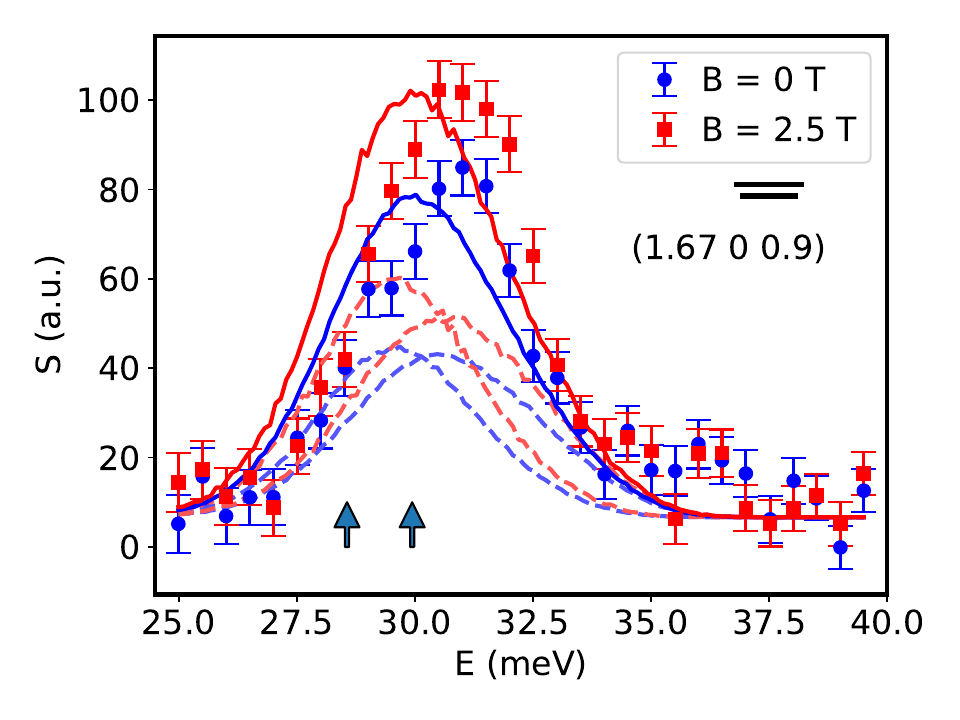} };
        \draw (-2.2, 2.2) node { \bf \fontfamily{phv} \selectfont (f) };
    \end{tikzpicture}
    \caption{Field dependence of the $\mathbf{Q} = \left(1.67,\ 0,\ l\right)$ magnon dispersion of MnTe.
    Panel (a) shows the theoretical dispersion from the linear spin-wave model, which we based on \cite{Liu2024},
    at zero field (blue curves) and at $B = 2.5\ \mathrm{T}$ (red curves).
    The individual scan positions along the $l$ direction are indicated as gray vertical bars.
    Panels (b)-(f) show the scan data for $0\ \mathrm{T}$ and $2.5\ \mathrm{T}$ (points)
    and instrumental resolution-convolutions of the theory (lines).
    The dashed blue (red) lines show the convoluted contributions of the two individual magnon modes on the 
    total intensity, which is depicted as a solid blue (red) line.
    For the resolution-convolution at $0\ \mathrm{T}$, the magnetic domains are averaged.
    The arrows show the theoretical magnon energies without correcting for resolution effects.
    The black horizontal lines in the upper right corners indicate the incoherent and the coherent full-width at half-maximum resolution of the instrument.
    The spin splitting at central $l$ positions of the branches
    as well as the increase of spin-spin correlation $S_{\perp}$ for increasing field are clearly 
    observed as apparent line broadening and as higher intensity, respectively.}
    \label{fig:exp_167}
    \end{centering}
\end{figure*}
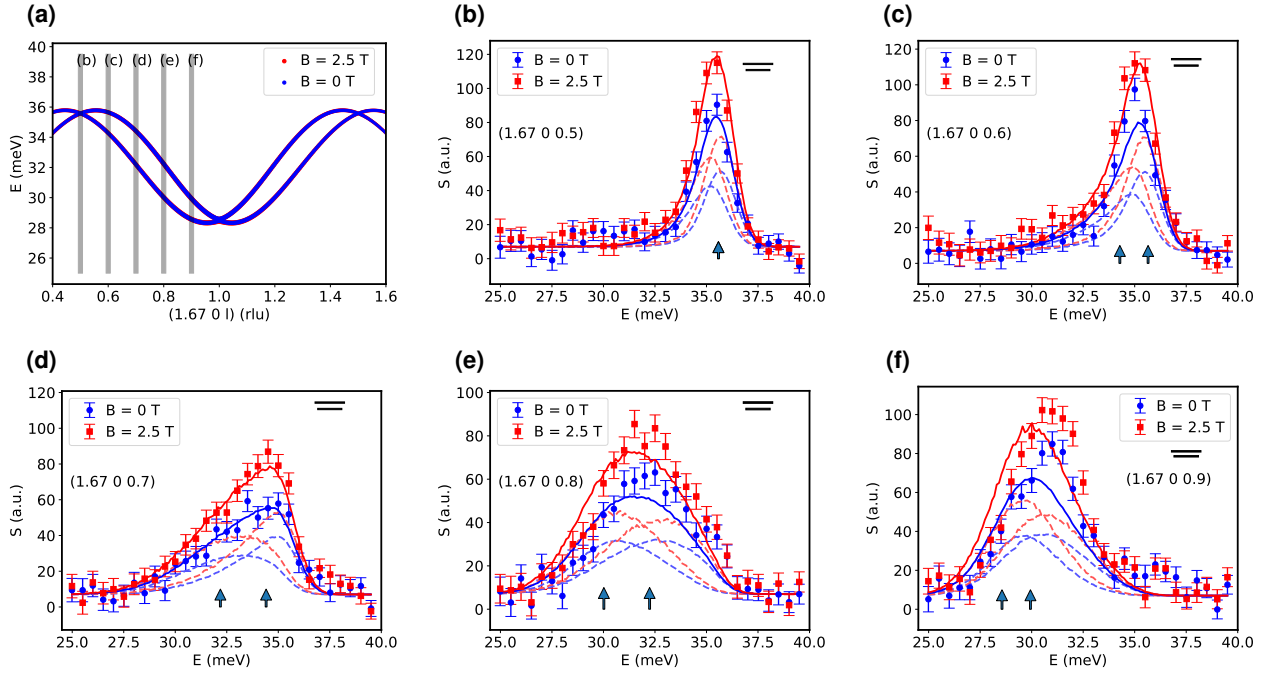

To prepare for the inelastic neutron scattering experiments, 
several dozens of small MnTe crystals having a total mass of approximately $1\ \mathrm{g}$ were co-aligned. 
For the alignment we used both X-ray diffraction and neutron Laue diffraction, 
the latter at the \textit{OrientExpress} \cite{OrientExpress} instrument.
The hexagonal crystallites were mounted on the front and back sides of two aluminum plates,
so that their crystallographic $c$ vectors were perpendicular to the plates' surfaces, see Fig.~\ref{fig:sample}.
We determined an effective mosaic spread of the co-aligned crystals of approximately $2.6^{\circ}$ full-width at half-maximum.

We measured the field dependence in MnTe at the thermal neutron triple-axis spectrometer \textit{IN8} \cite{IN8}
for several magnon dispersion branches that are accessible in the $\left(h0l\right)$ plane for fields up to
$B = 10\ \mathrm{T}$ along the reciprocal $\left(1\bar{2}0\right)$ direction,
which corresponds to the $-b$ direction in real space.
The field direction is perpendicular to the $\left(h0l\right)$ scattering plane.
For neutron wavelength selection, we used the $\left(200\right)$ reflection of a copper monochromator and the $\left(002\right)$ reflection of a graphite analyzer crystal.
To suppress possible phonons and other contamination, 
we cooled the sample to $T = 1.5\ \mathrm{K}$ under zero field and subtracted purely non-magnetic signals measured at room temperature.
The instrument was operated at a constant final neutron wave number of $k_f = 2.662\ \textup{\AA}^{-1}$, corresponding to an energy of $E_f = 14.68 \ \mathrm{meV}$.
In order to suppress higher order scattering, two graphite filters were installed on the secondary spectrometer.

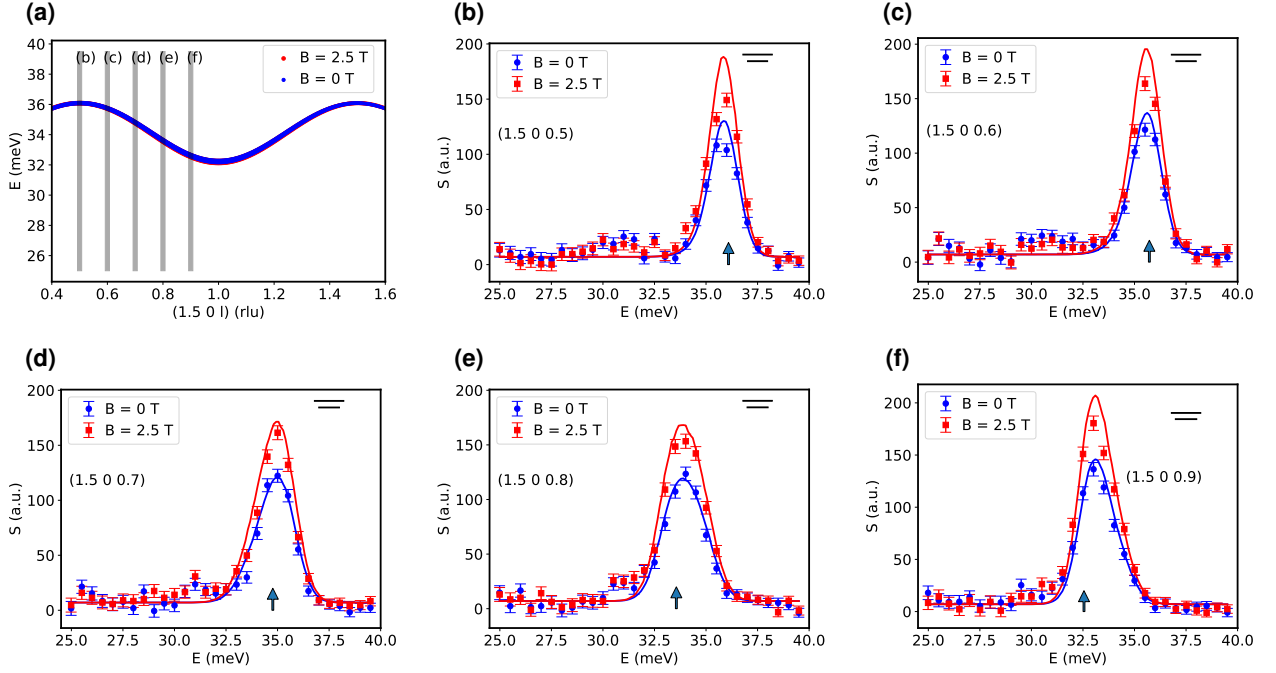
\begin{figure*}[htb]
    \begin{centering}
    \begin{tikzpicture}
        \draw (0, 0) node [ inner sep = 0 ] { \includegraphics[width=0.31\textwidth]{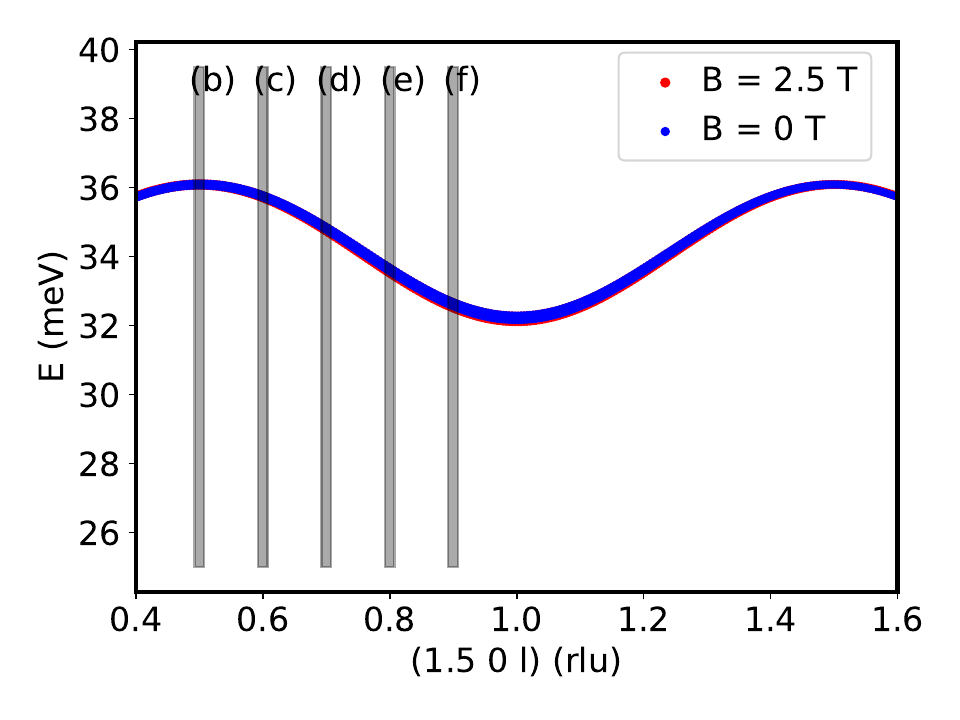} };
        \draw (-2.2, 2.2) node { \bf \fontfamily{phv} \selectfont (a) };
    \end{tikzpicture}
    \begin{tikzpicture}
        \draw (0, 0) node [ inner sep = 0 ] { \includegraphics[width=0.31\textwidth]{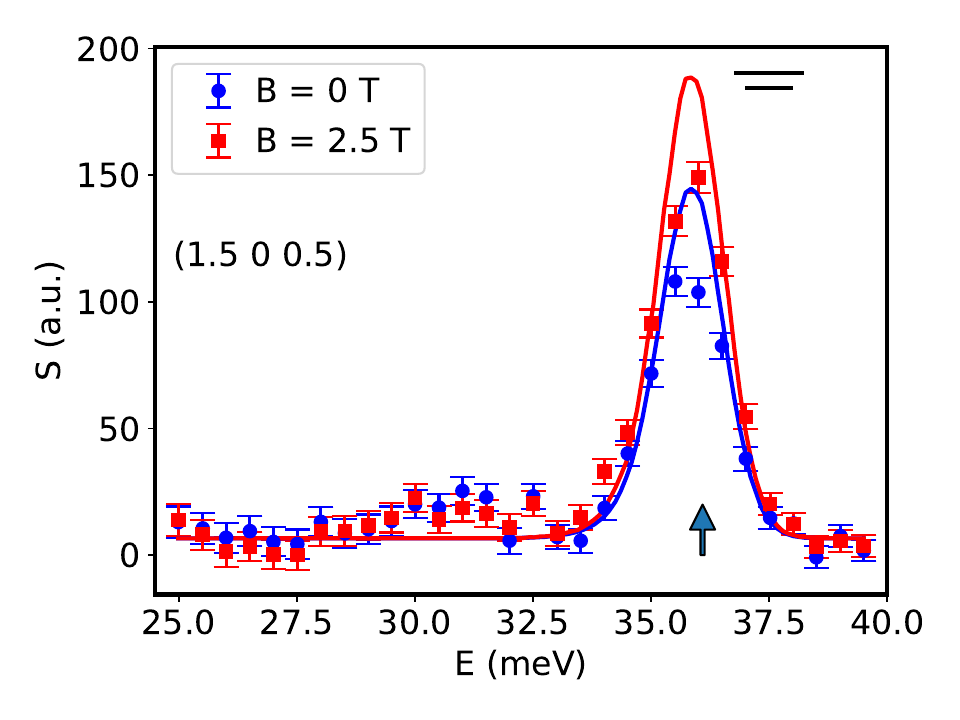} };
        \draw (-2.2, 2.2) node { \bf \fontfamily{phv} \selectfont (b) };
    \end{tikzpicture}
    \begin{tikzpicture}
        \draw (0, 0) node [ inner sep = 0 ] { \includegraphics[width=0.31\textwidth]{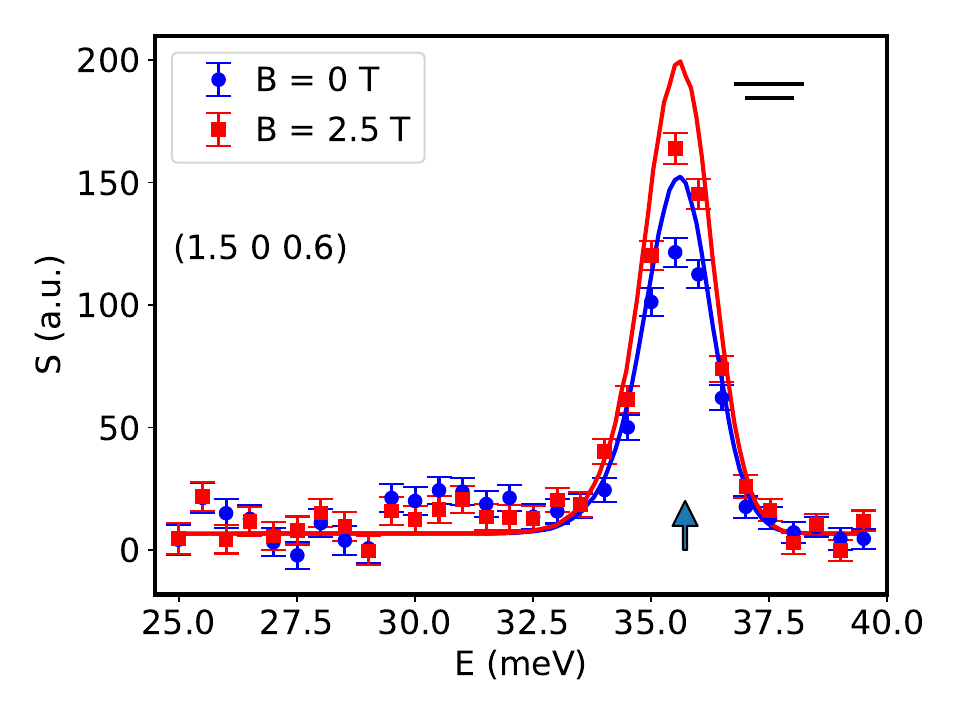} };
        \draw (-2.2, 2.2) node { \bf \fontfamily{phv} \selectfont (c) };
    \end{tikzpicture}
    \begin{tikzpicture}
        \draw (0, 0) node [ inner sep = 0 ] { \includegraphics[width=0.31\textwidth]{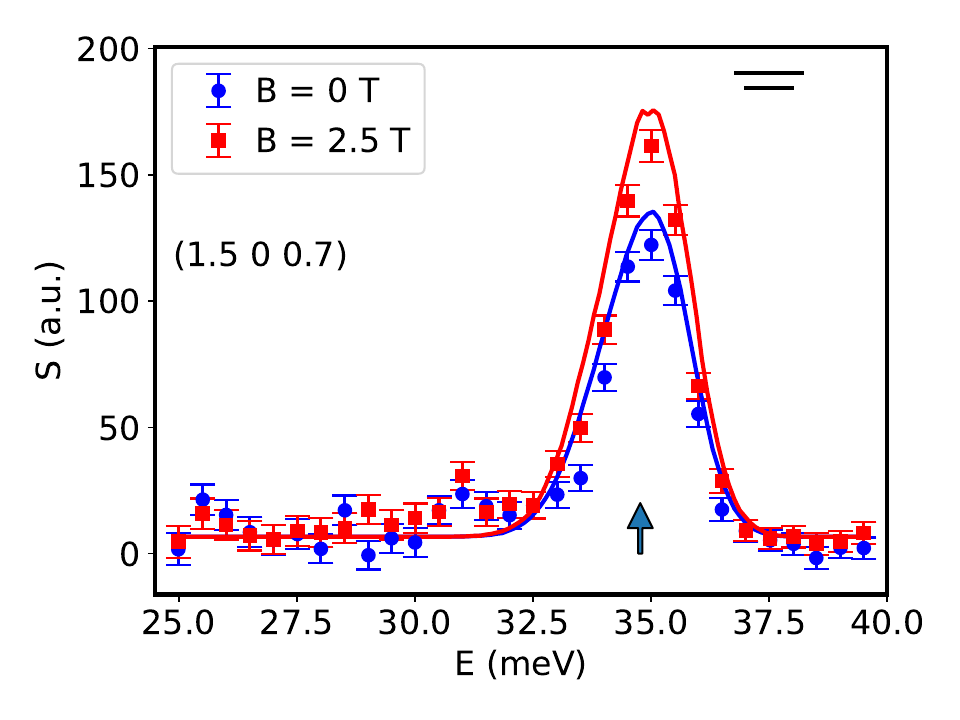} };
        \draw (-2.2, 2.2) node { \bf \fontfamily{phv} \selectfont (d) };
    \end{tikzpicture}
    \begin{tikzpicture}
        \draw (0, 0) node [ inner sep = 0 ] { \includegraphics[width=0.31\textwidth]{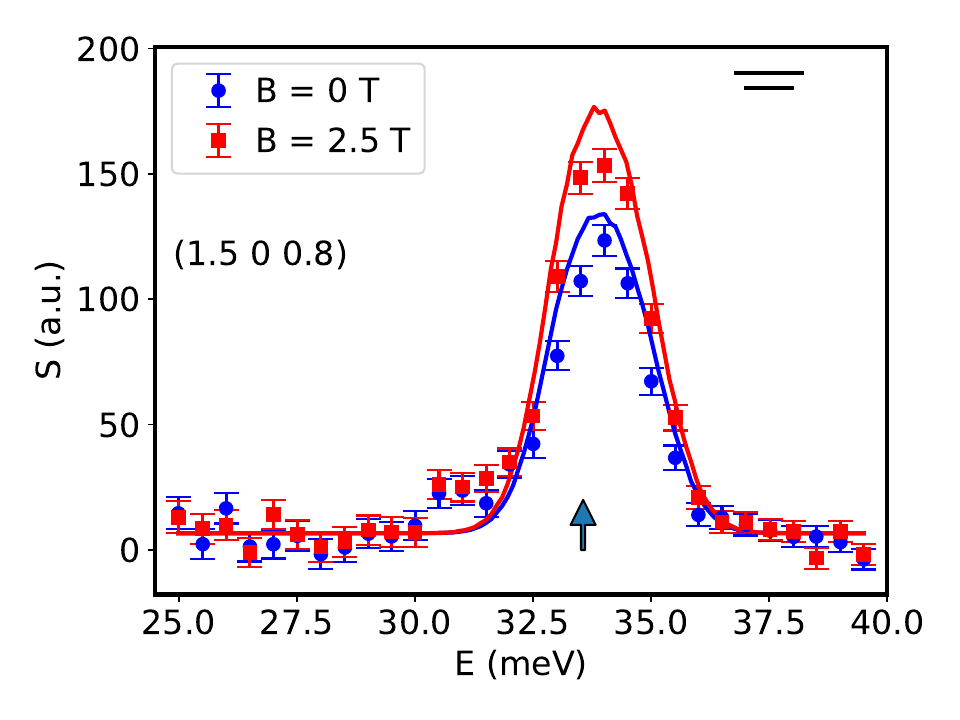} };
        \draw (-2.2, 2.2) node { \bf \fontfamily{phv} \selectfont (e) };
    \end{tikzpicture}
    \begin{tikzpicture}
        \draw (0, 0) node [ inner sep = 0 ] { \includegraphics[width=0.31\textwidth]{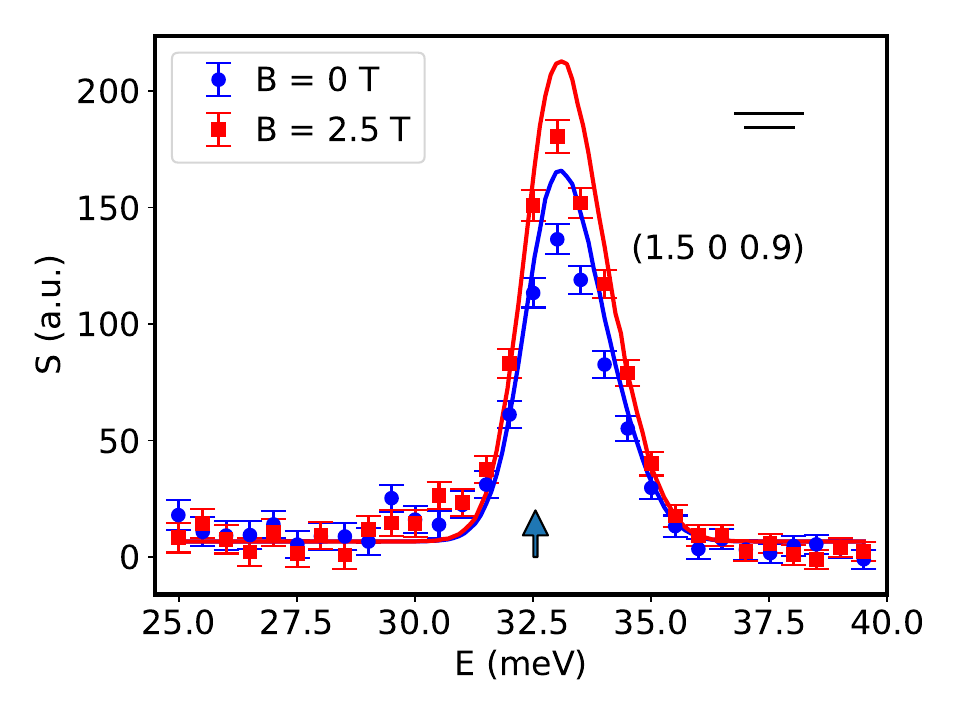} };
        \draw (-2.2, 2.2) node { \bf \fontfamily{phv} \selectfont (f) };
    \end{tikzpicture}
    \caption{Field dependence of the $\mathbf{Q} = \left(1.5,\ 0,\ l\right)$ magnon dispersion of MnTe.
    Panel (a) shows the theoretical dispersion from the linear spin-wave model, which we based on \cite{Liu2024},
    at zero field (blue curves) and at $B = 2.5\ \mathrm{T}$ (red curves).
    The individual scan positions along the $l$ direction are indicated as gray vertical bars.
    Panels (b)-(f) show the scan data for $0\ \mathrm{T}$ and $2.5\ \mathrm{T}$ (points)
    and instrumental resolution-convolutions of the theory (lines).
    For the resolution-convolution at $0\ \mathrm{T}$, the magnetic domains are averaged.
    The arrows show the theoretical magnon energies without correcting for resolution effects.
    The black horizontal lines in the upper right corners indicate the incoherent and the coherent full-width at half-maximum resolution of the instrument.
    The increase of spin-spin correlation $S_{\perp}$ for increasing field is clearly observed as higher intensity.
    No splitting of the magnon modes is observed.}
    \label{fig:exp_15}
    \end{centering}
\end{figure*}

The experiments could well reproduce both the splitting of the magnon modes at zero applied field, see Fig.~\ref{fig:exp_167},
which is most pronounced for the $\left(1.67,\ 0,\ l\right)$ dispersion branch,
as well as the absence of a splitting at $\left(1.5,\ 0,\ l\right)$, see Fig.~\ref{fig:exp_15}.
The splitting of the magnons along $\left(1.67,\ 0,\ l\right)$ can be observed indirectly by a strong line-broadening.
For both cases, we could register an increase of spectral weight on the magnons for increasing magnetic fields,
which is shown for finer-grained steps in applied field in Fig.~\ref{fig:exp_fields} (a).
Here, the intensity saturates for applied fields above approximately $1.5\ \mathrm{T}$, 
leading to an increase of the observed magnon intensity by up to 20\%. 
The origin of this effect is discussed in Sec.\ \ref{sec:results}.

The $\left(1.67,\ 0,\ l\right)$ and $\left(1.5,\ 0,\ l\right)$ dispersion branches are symmetrically equivalent to the $\left(\overline{1.33},\ 0,\ l \right)$ and $\left(\overline{1.5},\ 0,\ l \right)$ positions, respectively, measured by Liu \textit{et al.} \cite{Liu2024}.
Our results are shown in panels (b)-(f) of Figs. \ref{fig:exp_167} and \ref{fig:exp_15}, 
where the points show the experimental data and the lines depict a convolution of the instrumental resolution function \cite{Popovici1975, Takin2023} and the theoretical magnon model \cite{Liu2024}, 
which we calculated in our in-house software systems \cite{Weber2026Magpie, Magpie, Takin2023}, with only a global intensity scaling parameter as free variable to convert the intensities from theory to experiment.
The convolution integral was calculated via Monte-Carlo integration, 
which randomly selects points in the four-dimensional Gaussian
$\left(\mathbf{Q}, E\right)$ distribution function of the instrument's resolution.
For the resolution-convolution simulation at zero field,
the spin-spin correlation functions of the six magnetic domains,
whose N\'eel vectors point along
$\mathbf{L}_1 = \left(0 1 0\right)$,
$\mathbf{L}_2 = \left(0 \bar{1} 0\right)$,
$\mathbf{L}_3 = \left(1 0 0\right)$,
$\mathbf{L}_4 = \left(\bar{1} 0 0\right)$,
$\mathbf{L}_5 = \left(1 \bar{1} 0\right)$, and
$\mathbf{L}_6 = \left(\bar{1} 1 0\right)$, were averaged.
In real space, these N\'eel vectors correspond to the easy axes with the spins
pointing from one Mn to another Mn atom \cite{Kriegner2016}. 
For this work we assumed an equal population of all domains, which best matches the data.

The resolution-convolutions of the individual, spin-split magnon branches are shown as dashed lines in Fig.~\ref{fig:exp_167} (b-f).
Compared to the peak positions in the bare, unconvoluted linear spin-wave theory,
which are marked by arrows, we find a significant shift in energy.
Within the experimental resolution, the magnon excitations show no detectable intrinsic broadening. 
The calculated spectra reproduce the data best when only a small,
field-independent Gaussian broadening is introduced. 
Thus, linear spin-wave theory provides an adequate description of the measurements
without incorporating self-energy-induced line width corrections.

\section{Results\label{sec:results}}

\begin{figure}
    \begin{centering}
    \begin{tikzpicture}
        \hspace{0.1cm}
        \draw (0, 0) node [ inner sep = 0 ] { \includegraphics[width=0.92\columnwidth]{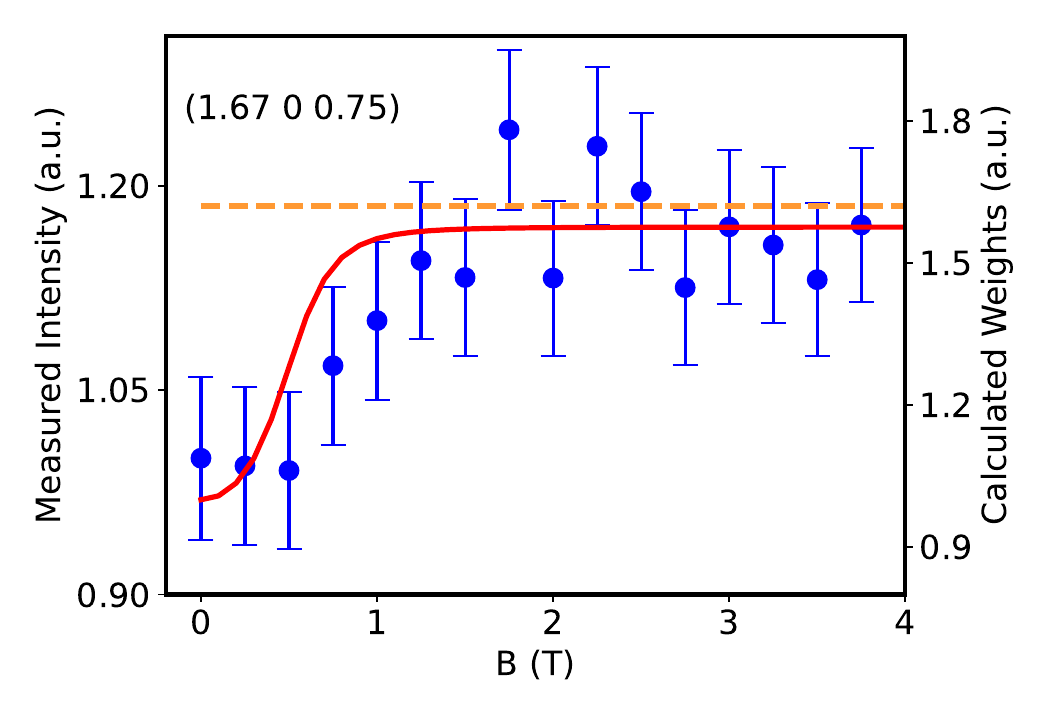} };
        \draw (1.25, -0.76) node [ inner sep = 0 ] { \includegraphics[width=0.37\columnwidth]{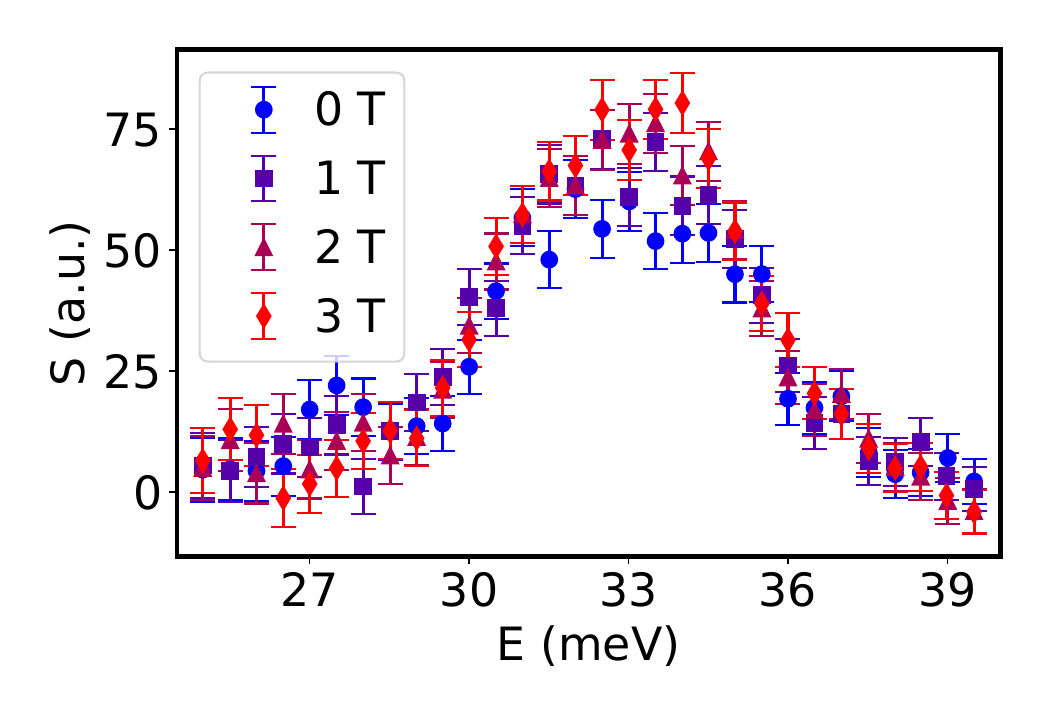} };
        \draw (-3.25, 2.4) node { \bf \fontfamily{phv} \selectfont (a) };
    \end{tikzpicture}
    \begin{tikzpicture}
        \draw (0, 0) node [ inner sep = 0 ] { 
        \hspace{-1.1cm}
        \includegraphics[width=0.78\columnwidth]{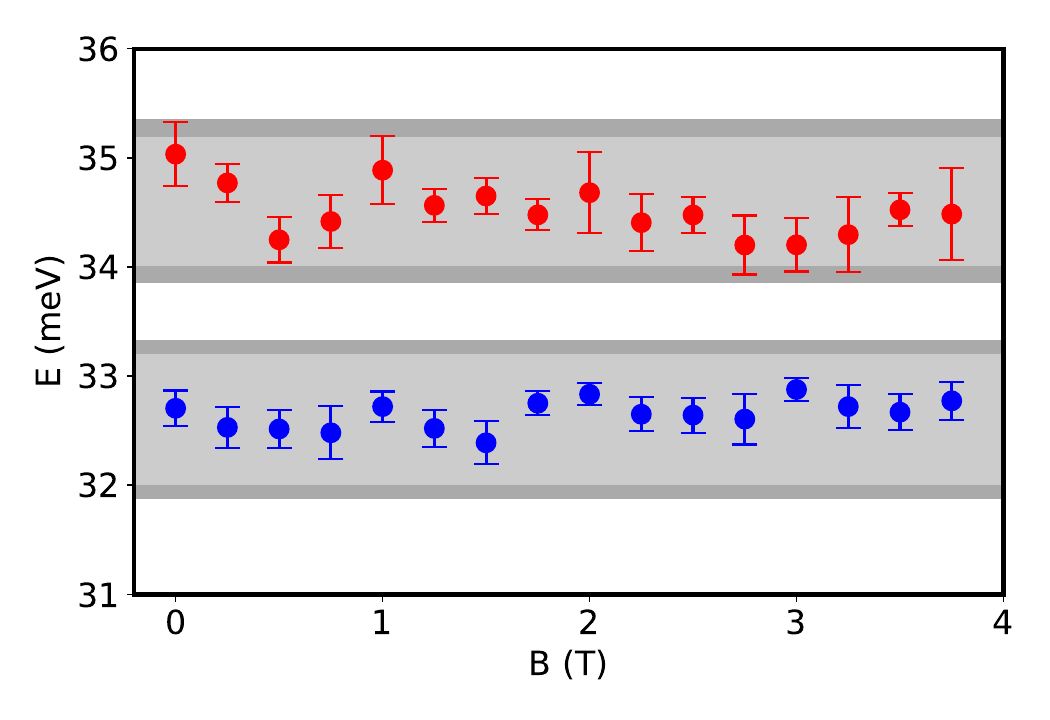} };
        \draw (-3.3, 2.4) node { \bf \fontfamily{phv} \selectfont (b) };
    \end{tikzpicture}
    \begin{tikzpicture}
        \draw (0, 0) node [ inner sep = 0 ] { 
        \hspace{-0.5cm}
        \includegraphics[width=0.795\columnwidth]{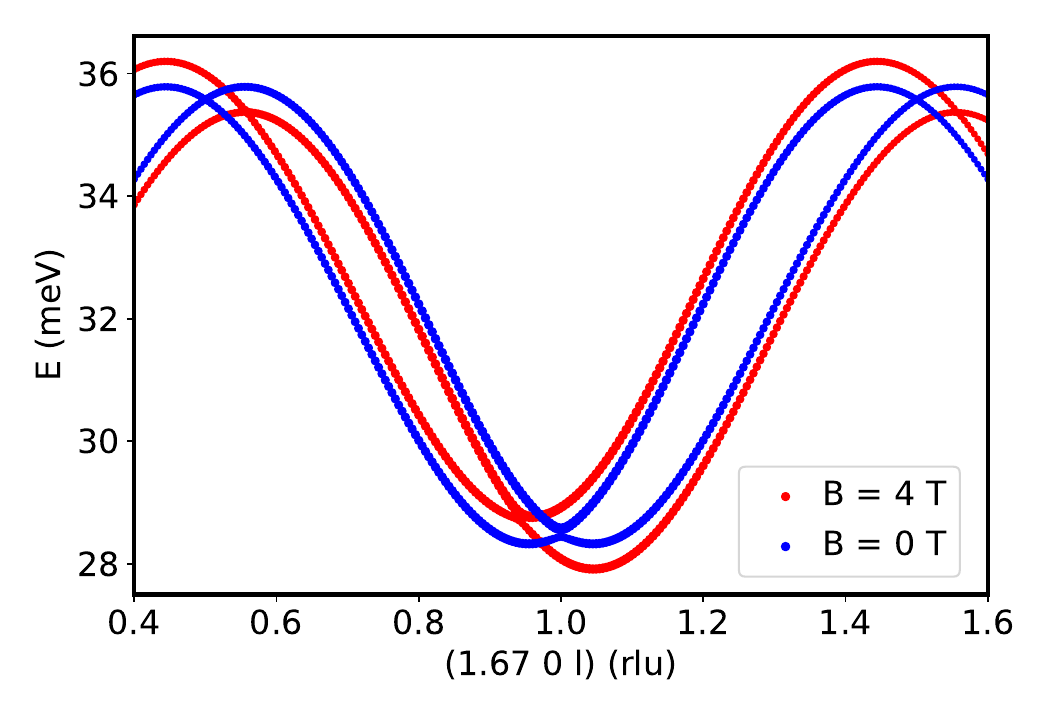} };
        \draw (-3.12, 2.4) node { \bf \fontfamily{phv} \selectfont (c) };
    \end{tikzpicture}
    \caption{(a) Integrated intensity of the experimentally measured $\mathbf{Q} = \left(1.67,\ 0,\ 0.75\right)$ magnon peaks vs. applied magnetic field (blue points)
    and calculated theoretical weights of the magnon modes including the orthogonal projector (solid red line), as well as without projector (dashed orange line). 
    The intensity saturates at about $1.5\ \mathrm{T}$.
    The inset shows constant-$Q$ scans at selected fields.
    (b) Fitted energies of the two magnon modes. The light (dark) gray regions mark the coherent (incoherent) full-width at half-maximum energy resolution of the instrument.
    (c) Comparison of the sole effect of the field on the dispersion without adapting the ground state.}
    \label{fig:exp_fields}
    \end{centering}
\end{figure}

We observe spin-split magnon modes in a co-aligned assembly of approximately 50 MnTe crystallites. 
The integrated intensities of both the chiral and degenerate magnon modes increase with applied magnetic field, whereas their energies and line widths remain unchanged within experimental resolution. 
At an energy transfer of $34\ \mathrm{meV}$, the coherent and incoherent energy resolutions are approximately $\Delta E_{c} = 1.21\ \mathrm{meV}$ and $\Delta E_{i} = 1.47\ \mathrm{meV}$, respectively, quoted as full-widths at half-maximum. 
The observed line widths are therefore resolution-limited, with the broadening arising primarily from the large mosaic spread of the co-aligned crystallites.
The field-dependent intensity effect is illustrated in Fig.~\ref{fig:exp_fields} (a) for the modes at $\mathbf{Q} = \left(1.67,\ 0,\ 0.75\right)$.
The stability of their energies within the instrumental resolution is shown in Fig.~\ref{fig:exp_fields} (b).

The field-independent magnon dispersion is consistent with a continuous reorientation of the N\'eel vector.
For a fixed N\'eel vector, our calculations predict measurable field-induced changes in the dispersion even for fields small compared with the dominant exchange scale. 
This is illustrated in Fig.~\ref{fig:exp_fields} (c) where we compare the spectra obtained from linear spin-wave theory along the
$\left( 1.67,\ 0,\ l \right)$ dispersion branch for zero and $4$ T, assuming a fixed N\'eel vector
$\mathbf{L} = \left( 010 \right)$. 
In contrast to our experimental observations, the spectra differ noticeably for the two field strengths,
both with respect to the observed stability of the energies and the symmetry of the dispersion.

The field-induced increase in magnon intensity can be explained by the reorientation of the magnetic ground state as well. 
Inelastic neutron scattering probes the component of the dynamical structure factor transverse to the momentum transfer and is therefore sensitive to the orientation of $\mathbf{L}$ through the projector in Eq.~\eqref{equ:Sperp}. 
Fig.~\ref{fig:exp_fields} (a) shows the integrated intensity at $\mathbf{Q} = \left( 1.67, 0, 0.75 \right)$ as a function of field strength. 
The data agree closely with the calculated projected spectral weight, shown by the solid red line. 
By contrast, the non-projected dynamical structure factor $S_{xy}(\mathbf{Q}, E)$, shown by the dashed orange line, remains constant over the experimentally accessible field range. 
The observed field dependence therefore originates from the changing transverse projection rather than from a modification of the underlying spin dynamics.
The magnetic ground state orientation can thus be controlled by field without measurably altering the magnon spin splitting, energies, or lifetimes.

This further suggests that the relevant energy scale of the Dzyaloshinskii-Moriya interactions,
which control the stiffness of the N\'eel-vector, are indeed well separated from that of the exchange
interactions.
Although spin-orbit coupling substantially modifies the electronic band structure of MnTe \cite{krempasky2024altermagnetic}, we find Dzyaloshinskii-Moriya interactions to be negligible for the measured magnon response. 
Despite the global centrosymmetry of the crystal, antisymmetric exchange is symmetry-allowed on the relevant Mn--Mn bond because the bond midpoint is not an inversion center \cite{moriya}. 
Indeed, reversing the external magnetic field produces no resolvable change in the spectra ruling out any non-reciprocal effects \cite{Weber2026}. 
The observed peak positions are instead strongly affected by instrumental resolution, particularly at $\mathbf{Q} = \left(1.67, 0, 0.8 \right)$ and $\mathbf{Q} = \left(1.67, 0, 0.9 \right)$ as shown in Fig.~\ref{fig:exp_167} (e, f). 
Resolution convolution is therefore essential for reproducing the apparent shifts.

\section{Discussion and Conclusion}
The absence of detectable changes in the shapes of the dispersion branches and their line widths demonstrates that experimentally accessible fields reorient the altermagnetic order without appreciably perturbing the exchange-dominated magnon excitations.
Together with the pronounced field dependence of the measured spectral weight, our results reveal a clear separation between the soft orientational degree of freedom of the N\'eel vector and the robust chiral magnon spectrum. 
The magnetic field therefore controls how the excitations couple to an external probe while leaving their frequencies, spin splitting, and coherence essentially unchanged.

Our conclusion of a field-dependent rotation of the N\'eel vector is consistent with
magnetoresistance measurements in MnTe done by Kriegner \textit{et al.} \cite{Kriegner2016}. 
Kriegner \textit{et al.} furthermore report a spin-flop transition with the spins canting towards the
magnetic field direction at field magnitudes between 2 and 5 T.
Such a canted ground state would modify the magnon spectrum.
However, our spin-wave calculations show that at the low angles of spin canting that
Kriegner \textit{et al.} report, namely $1^{\circ}$,
this modification is too subtle to be detected in triple-axis measurements.

The observed separation of energies is attractive for the research of reconfigurable magnonic devices. 
A future implementation would be a fixed-frequency magnonic logic gate or router in which rotation of the N\'eel vector changes the coupling of a spin-polarized injector or detector to the two chiral magnon branches. The transmitted amplitude or polarization channel could then be switched without de-tuning the magnon band, modifying its group velocity, or compromising its line width. This is particularly useful in resonant or interference-based circuits, where changes in the dispersion would otherwise disrupt phase matching and require retuning of downstream components.
Within a simple ballistic-flux picture based on the semiclassical magnon Boltzmann transport,
the contribution of each mode to the incident flux is weighted by its group velocity normal
to the interface and by the corresponding matrix element \cite{ballistic}.
A reduction of the velocity-weighted average spectral weight would therefore reduce the coupled magnon flux.
However, because the spectral weight change as well as the matrix elements are in general nonuniform
across momentum space, the observed suppression of roughly 20\% does not in general translate directly
into a 20\% reduction of the flux.
Theoretical work has indeed predicted that reorienting the N\'eel vector can switch or redirect
transverse magnon currents in altermagnets \cite{lei2026vectormagnonicselectricalinjection}.

MnTe therefore emerges as a promising platform for controlling the coupling to altermagnetic magnons without modifying their exchange-defined spectrum.
Because the bulk $g$-wave symmetry of MnTe can suppress a net magnon spin current, transport devices may require additional symmetry breaking through strain, interfaces, or patterned hetero-structures.
Recent demonstrations of strain-controlled N\'eel-vector rotation, together with proposals for strain-induced changes of the magnetic symmetry, suggest a practical route toward this functionality \cite{straingiant, Liu2026}. 
Our results indicate that such control could be implemented while preserving the characteristic magnon splitting and resolution-limited line widths.

\section*{Acknowledgments}
Thanks to Franck Charpenay, Philippe Decarpentrie, Andrew Wildes, and O. Fabelo Rosa (all at ILL) for technical support.
We wish to thank Alexander Mook (University of M\"unster), Elisa Rebolini (ILL), Marie-Bernadette Lepetit (ILL), and Henrik Jacobsen (ESS DMSC) for theory discussions. 
We thank Shirin Mozaffari (Clemson University) as well as Matthew Cothrine, David Mandrus, and Md Serajus Salekin Chowdhury (all at the University of Tennessee) for providing backup MnTe crystals.

N.H. acknowledges financial support through the Feodor-Lynen fellowship awarded by the Humboldt Foundation.
S.S. and N.H.J. acknowledge support from the National Science Foundation (NSF) through CAREER grant (Award No. DMR-2337535) and the Materials Research Science and Engineering Center at the University of Michigan (Award No. DMR-2309029).

\section*{Author Contributions}
The study was proposed by T.W., N.H., and D.K. 
The neutron scattering experiments were conducted by T.W. The powder X-ray diffraction experiment was conducted by S.S.
A.I. and A.P. provided support at the \textit{IN8} instrument.
The samples were grown, characterized, and provided by S.S., R.T., A.H., A.K.M.A.S, T.P.J., and N.H.J.
The sample crystals were co-aligned by T.W.
Data analysis was done by T.W.
T.W. and N.H. performed theoretical calculations.
The manuscript was written by T.W., N.H., S.S., N.H.J, S.B., and D.K. with additional contributions from V.M. and S.K.
All authors discussed the manuscript.

\section*{Data Availability}
The experiments were performed at the Institut Laue-Langevin (ILL) in 2025 and 2026 using the \textit{IN8} spectrometer \cite{IN8} and have the DOIs \href{https://doi.org/10.5291/ILL-DATA.4-01-1868}{10.5291/ILL-DATA.4-01-1868} 
and \href{https://doi.org/10.5291/ILL-DATA.4-01-1911}{10.5291/ILL-DATA.4-01-1911}, respectively.

The data and code used for data analysis is available under DOI \href{https://doi.org/10.5281/zenodo.20623007}{10.5281/zenodo.20623007} \cite{DataAnalysis}.
The software used for data analysis is available as open source \cite{Weber2026Magpie, Magpie, Takin2023}.

%

\end{document}